\documentclass[prb,showpacs,twocolumn]{revtex4}
\usepackage{amssymb}
\usepackage{amsmath}
\usepackage{bm}
\usepackage{graphicx}
\usepackage{dcolumn}

\begin{document}

\title{Angular dependence of magnetoresistance in strongly anisotropic
quasi-two-dimensional metals for various Landau-level shapes}
\author{T.I.~Mogilyuk}
\address{National Research Centre "Kurchatov Institute", Moscow, Russia}

\author{P.D.~Grigoriev} \email[Corresponding author; e-mail:
]{grigorev@itp.ac.ru}

\address{L. D. Landau Institute for Theoretical Physics, 142432
Chernogolovka, Russia}

\begin{abstract}
We present the quantum-mechanical calculations of the angular
dependence of interlayer conductivity $\sigma _{zz}(\theta )$ in a
tilted magnetic field in quasi-2D layered metals. Our calculation
shows that the LL shape is important for this angular dependence.
In particular, the amplitude of angular magnetoresistance
oscillations (AMRO) is much stronger for the Gaussian LL shape
than for the Lorentzian. The ratio $\sigma _{zz}\left( \theta
=0\right) /\sigma _{zz}\left( \theta \rightarrow \pm 90^{\circ
}\right) $ is also several times larger for the Gaussian LL shape.
AMRO and Zeeman energy splitting lead to a spin current. For
typical organic metals and for a medium magnetic field 10T this
spin current is only a few percent of the charge current, but its
value may almost reach the charge current for special tilt angles
of magnetic field. The spin current has strong angular
oscillations, which are phase-shifted as compared to the usual
AMRO.
\end{abstract}

\date{\today }
 \maketitle

\section{Introduction}

The angular magnetoresistance oscillations (AMRO) is a prominent feature of
strongly anisotropic layered conductors, which gives an important
information about their electronic properties (see, e.g., \cite%
{MarkReview2004,Singleton2000Review,KartPeschReview} for reviews). \ AMRO
are actively used to investigate various layered compounds, including
organic metals,\cite%
{MarkReview2004,Singleton2000Review,KartPeschReview,OMRev,MQORev,Brooks2006,LebedBook}
high-temperature cuprate superconductors,\cite%
{HusseyNature2003,AbdelNature2006,AbdelPRL2007AMRO,McKenzie2007,HelmNd2010}
heterostructures\cite{Kuraguchi2003} etc.

AMRO were first observed\cite{KartsAMRO1988} in 1988 in a quasi-2D strongly
anisotropic organic metal $\beta $-(BEDT-TTF)$_{2}$IBr$_{2}$. The first
explanation of AMRO appeared next year\cite{Yam} and used the geometrical
arguments for the Fermi surface of the corrugated-cylinder shape, which
corresponds to strongly anisotropic electron dispersion%
\begin{equation}
\epsilon _{3D}\left( \mathbf{k}\right) \approx \epsilon _{2D}\left(
k_{x},k_{y}\right) -2t_{z}\cos (k_{z}d),  \label{ES3D}
\end{equation}%
where $\hbar k_{z}$ is out-of-plane electron momentum, $\hbar $ is the
Planck's constant, $d$ is the interlayer spacing, and the interlayer
transfer integral $t_{z}$ is much less than the Fermi energy $E_{F}$. For
the quadratic and isotropic in-plane electron dispersion $\epsilon
_{2D}\left( k_{x},k_{y}\right) =\hbar ^{2}\left( k_{x}^{2}+k_{y}^{2}\right)
/2m^{\ast }$ Yamaji obtained\cite{Yam} that the minima of interlayer
conductivity $\sigma _{zz}\left( \theta \right) $ correspond to the zeros of
$J_{0}\left( \kappa \right) $, where $J_{0}$ is the Bessel function of the
zeroth order, $\kappa \equiv k_{F}d\tan \theta $, $k_{F}$ is the in-plane
Fermi momentum, and $\theta $ is the angle between the applied magnetic
field $\boldsymbol{B}$ and the normal to the conducting planes. The direct
calculation of interlayer conductivity, using the electron dispersion in Eq.
(\ref{ES3D}) and the Boltzman transport equation in the $\tau $%
-approximation, gives\cite{Yagi1990}%
\begin{equation}
\sigma _{zz}=\sigma _{zz}^{0}\left\{ \left[ J_{0}\left( \kappa \right) %
\right] ^{2}+2\sum_{\nu =1}^{\infty }\frac{\left[ J_{\nu }\left( \kappa
\right) \right] ^{2}}{1+\left( \nu \omega _{c}\tau \right)^2 }\right\} ,
\label{sa}
\end{equation}%
where the cyclotron frequency $\omega _{c}=eB_{z}/m^{\ast }c$, $\tau $ is
the mean free time, and the interlayer conductivity without magnetic field%
\begin{equation}
\sigma _{zz}^{0}=e^{2}\rho _{F}\left\langle v_{z}^{2}\right\rangle \tau
=2e^{2}t_{z}^{2}m^{\ast }\tau d/\pi \hbar ^{4},  \label{s0}
\end{equation}%
where $\rho _{F}=m^{\ast }/\pi \hbar ^{2}d$ is the 3D DoS at the Fermi level
in the absence of magnetic field per two spin components, and the mean
square interlayer electron velocity $\left\langle v_{z}^{2}\right\rangle
=2t_{z}^{2}d^{2}/\hbar ^{2}$. Here $e$ is the electron charge, $m^{\ast }$
is the effective electron mass, $B_{z}$ is the component of magnetic field
perpendicular to conducting layers, and $c$ is the light velocity.\ Eq. (\ref%
{sa}) agrees with the result of Yamaji at $\omega _{c}\tau \rightarrow
\infty $. A microscopic calculation\cite{Kur} of quasi-2D AMRO also gives
Eq. (\ref{sa}) when the number of filled Landau levels (LLs) $n_{LL}^{F}\gg 1
$.

The calculations of AMRO in Refs. \cite%
{Yam,Yagi1990,Kur,Mark92,Singleton2001,GrigAMRO2010,ShofieldCooper2000}
assume a well-defined 3D electron dispersion (\ref{ES3D}), i.e. that the LL
separation $\hbar \omega _{c}$ and broadening $\Gamma _{0}=\hbar /2\tau $
are much less than $t_{z}$. The inverse "weakly incoherent" limit $t_{z}\ll
\Gamma _{0}$ with the momentum-conserving "coherent" interlayer hopping was
also considered.\cite{MosesMcKenzie1999} The interlayer conductivity was
calculated as a tunnelling conductivity between two adjacent conducting
layers with short-range disorder in magnetic field, which again resulted to
Eq. (\ref{sa}).\cite{MosesMcKenzie1999} The calculation in Refs. \cite%
{MosesMcKenzie1999,Kur} is performed under the assumption that the electron
self-energy $|\text{Im}\Sigma| =\Gamma _{0}$ is independent of energy and magnetic
field. This assumption, being almost equivalent to the $\tau $%
-approximation, is incorrect in 2D or strongly anisotropic quasi-2D layered
compounds with $t_{z},\Gamma _{0}\lesssim \hbar \omega _{c}$, i.e. in the
presence of strong magnetic quantum oscillations (MQO).\cite%
{Ando1,Ando,ShubCondMat1,ChampelMineev,Shub,Gvozd2004,WIPRB2011,WIJETP2011,WIFNT2011,GrigPRB2013,WIPRB2012}
Even if MQO are suppressed by temperature or long-range disorder, that smear
the Fermi distribution function,\cite{Shoenberg,Abrik,Ziman} these MQO
produce the monotonic growth\cite%
{WIPRB2011,WIJETP2011,WIFNT2011,GrigPRB2013,WIPRB2012} of the longitudinal
interlayer magnetoresistance $R_{zz}\left( B_{z}\right) =1/\sigma _{zz}$ and
of the LL broadening $\Gamma =\Gamma \left( B_{z}\right) $, which changes
the angular dependence $R_{zz}\left( \theta \right) $.\cite{WIPRB2011} In
Ref. \cite{WIPRB2011} a simple amendment to Eq. (\ref{sa}) in the limit $%
t_{z}\ll \Gamma _{0}$ was proposed (see Eq. (36) of Ref. \cite{WIPRB2011}),
which includes the renormalization of the prefactor,
\begin{equation}
\sigma _{zz}^{0}\rightarrow \sigma _{zz}^{0}\left( B_{z}\right) \approx
\sigma _{zz}^{0}\left[ 1+\left( 2\omega _{c}\tau \right) ^{2}\right] ^{-1/4},
\label{s0R}
\end{equation}%
and the similar renormalization of the effective mean scattering time in Eq.
(\ref{sa}),
\begin{equation}
\tau \rightarrow \tau \left( B_{z}\right) \approx \tau \left[ 1+\left(
2\omega _{c}\tau \right) ^{2}\right] ^{-1/4}.  \label{tau0R}
\end{equation}%
However, this modification of Eq. (\ref{sa}) has also several drawbacks.
First, it disregards the additional "quantum" term, coming from MQO. This
term was first obtained in Refs. \cite{ShubCondMat1,ChampelMineev,Shub} for
magnetic field perpendicular to the conducting layers. For tilted magnetic
field this "quantum" term is given by the second term in the curly brackets
in Eq. (\ref{Kubo1}) below or in Eq. (29) of Ref. \cite{WIPRB2011}. Second,
Eq. (\ref{sa}) is derived for the Lorentzian Landau level (LL) shape while
the actual LL shape in strong magnetic field for strongly anisotropic
quasi-2D layered metals with $t_{z}\ll \hbar \omega _{c}$ is closer to
Gaussian.\cite{WIPRB2012} The aim of this paper is to perform a more
rigorous calculation of the angular dependence of interlayer
magnetoresistance in strongly anisotropic quasi-2D metals, which includes
the contribution from MQO and is applicable for various LL shapes.

Eq. (\ref{sa}) gives the dependence only on the polar tilt angle $\theta $
of magnetic field, because it assumes an isotropic in-layer dispersion $%
\epsilon _{2D}\left( k_{x},k_{y}\right) $. Its generalization for the
anisotropic in-plane dispersion also within the $\tau $-approximation was
considered analytically in Refs. \cite{Mark92,Singleton2001,GrigAMRO2010}
and numerically in Refs. \cite%
{HusseyNature2003,AbdelNature2006,AbdelPRL2007AMRO,McKenzie2007}, which also
gives the azimuthal-angle dependence of MR.

\section{Calculation}

To calculate the interlayer conductivity $\sigma _{zz}$, we use the same
two-layer model as in Ref. \cite{WIPRB2011,WIJETP2011,MosesMcKenzie1999}. We
consider only two adjacent conducting layers with short-range impurities and
with a coherent interlayer electron hopping, which conserves the in-plane
electron momentum (see Eqs. (8)-(11) of Ref. \cite{WIPRB2011}). This model
is applicable when the interlayer transfer integral $t_{z}$ is less the LL
broadening.\cite{GrigPRB2013} The interlayer conductivity can be calculated
using the Kubo formula, which is valid when many LL are filled, so that
there is no in-plane electron localization as in the quantum Hall effect. As
a result, after introducing the additional sum over spin orientation $s=\pm 1
$, one obtains (compare to Eq. (29) of Ref. \cite{WIPRB2011} and to Eq. (50)
of Ref. \cite{MosesMcKenzie1999}):
\begin{align}
& \sigma _{zz}=\frac{e^{2}t_{z}^{2}d}{\pi \hbar }\sum_{s=\pm 1}\int
d\varepsilon \left[ -n_{F}^{\prime }(\varepsilon )\right] \times
\label{Kubo1} \\
& \times \int d^{2}{\boldsymbol{r}}\left\{ \left\vert \left\langle G_{R}({%
\boldsymbol{r}},\varepsilon )\right\rangle \right\vert ^{2}\cos \left(
iqy\right) \right. -  \notag \\
& \left. \text{Re}\left[ \left\langle G_{R}({\boldsymbol{r}},\varepsilon
)\right\rangle \left\langle G_{R}(-{\boldsymbol{r}},\varepsilon
)\right\rangle e^{iqy}\right] \right\} .  \notag
\end{align}%
Here $n_{F}^{\prime }(\varepsilon )=-1/\{4T\cosh ^{2}\left[ (\varepsilon
-\mu )/2T\right] \}$ is the derivative of the Fermi distribution function, $%
\mu $ is the chemical potential, ${\boldsymbol{r\equiv }}\left\{ x,y\right\}
$, $q\equiv eBd\sin \theta /\hbar c$ and $\left\langle G_{R}({\boldsymbol{r}}%
,\varepsilon )\right\rangle =\left\langle G_{R}({\boldsymbol{r}}_{1},{%
\boldsymbol{r}}_{1}+{\boldsymbol{r}},\varepsilon )\right\rangle $ is the
retarded electron Green's function as function of the coordinate and energy,
averaged over impurity positions. The impurity averaging of each Green's
function in Eq. (\ref{Kubo1}) can be performed separately, because the
vertex corrections have the next order of smallness in the parameter $t_{z}$
and because the impurities are short-range. For short-range impurities in
the non-crossing approximation the averaged electron Green's function is
given by\cite{Ando,WIPRB2011}
\begin{equation}
G({\boldsymbol{r}}_{1},{\boldsymbol{r}}_{2},\varepsilon )=\sum_{n,k_{y}}\Psi
_{n,k_{y}}^{\ast }(r_{2})\Psi _{n,k_{y}}(r_{1})G\left( \varepsilon ,n\right)
,  \label{GNC}
\end{equation}%
where $n$ is the LL number,
\begin{equation}
G\left( \varepsilon ,n\right) =\frac{1}{\varepsilon -\varepsilon
_{2D}(n)-\Sigma \left( \varepsilon \right) },  \label{G1}
\end{equation}%
and $\epsilon _{2D}(n)\equiv \epsilon _{n}=\hbar \omega _{c}\left(
n+1/2\right) $. In the noncrossing approximation the electron self-energy $%
\Sigma \left( \varepsilon \right) $ depends only on energy $\varepsilon $
(see Appendix of Ref. \cite{WIPRB2011}), being a periodic function with the
period $\hbar \omega _{c}$. Moreover, Eq. (\ref{GNC}) contains the bare
electron wave functions%
\begin{equation}
\Psi _{n,k_{y}}(r)=\Psi _{n}(x-l_{H}^{2}k_{y})\exp (ik_{y}y),  \label{Psi}
\end{equation}%
where
\begin{equation}
\Psi _{n}\left( x\right) =\frac{\exp \left( -x^{2}/2l_{H}^{2}\right)
H_{n}\left( x/l_{H}\right) }{\left( \pi l_{H}^{2}\right) ^{1/4}2^{n/2}\sqrt{%
n!}},  \label{Gg}
\end{equation}%
$H_{n}\left( x/l_{H}\right) $ is the Hermite polynomial and $l_{H}=\sqrt{%
\hbar c/eB_{z}}$ is the magnetic length. This considerably simplifies the
calculation. For the retarded and advanced Green's functions the sign of Im$%
\Sigma $ is fixed. In strong magnetic field $\hbar \omega _{c}\gg \Gamma _{0}
$ the electron Green's function $G\left( \varepsilon ,n\right) $ can be
calculated restricting to only one LL at $\varepsilon \approx \varepsilon
_{2D}(n)$, which in the noncrossing approximation gives a dome-like rather
than Lorentzian LL shape.\cite{Ando} The inclusion of diagrams with the
intersection of impurity lines adds the exponential tails in the electron
density of states (DoS) $\rho \left( \varepsilon \right) =-$Im$G\left(
\varepsilon ,n\right) /\pi $ for each LL.\cite{Ando3} As the angular
dependence of MR depends on the LL shape, we first calculate Eq. (\ref{Kubo1}%
) without restriction to any particular form of the electron Green's
function, and then compare the results for various LL shapes.

Now we substitute the electron Green's function from Eq. (\ref{GNC}) to Eq. (%
\ref{Kubo1}). The first term in curly brackets, coinciding with Eq. (50) of
Ref. \cite{MosesMcKenzie1999} and responsible for the so-called "classical" $%
G_{R}G_{A}$ part of conductivity $\sigma _{zz}$, rewrites as (see Appexdix
I)
\begin{equation}
\begin{split}
& Cl\equiv \int d^{2}r|G(r,\epsilon )|^{2}\cos \left( qy\right) \\
& =g_{LL}\sum_{n,p\in Z}\left[ \text{Re}G(\epsilon ,n)\text{Re}G(\epsilon
,n+p)\right] + \\
& \left. \text{ Im}G(\epsilon ,n)\text{Im}G(\epsilon ,n+p)\right] Z(n,p),
\end{split}
\label{Cl}
\end{equation}%
where the LL degeneracy per unit area $g_{LL}=1/2\pi l_{H}^{2}=eB_{z}/2\pi
\hbar c$,
\begin{eqnarray}
Z(n,p) &=&\exp \left( -\frac{(ql_{H})^{2}}{2}\right) \left( \frac{%
(ql_{H})^{2}}{2}\right) ^{p}\times  \label{Z} \\
&&\left( L_{n}^{p}\left( \frac{(ql_{H})^{2}}{2}\right) \right) ^{2}\left(
\frac{n!}{(n+p)!}\right) ,  \notag
\end{eqnarray}%
and $L_{n}^{p}\left( x\right) $ is the Laguerre polynomial. The second term
in curly brackets in Eq. (\ref{Kubo1}), which is absent in Refs. \cite%
{Yagi1990,MosesMcKenzie1999} and responsible for the so-called "quantum"
part of conductivity $\sigma _{zz}$, rewrites as (see Appexdix II)

\begin{equation}
\begin{split}
& Q\equiv \text{Re}\left[ \int d^{2}r\left\langle G_{R}({\boldsymbol{r}}%
,\varepsilon )\right\rangle ^{2}\exp \left( iqy\right) \right] \\
& =g_{LL}\sum_{n,p\in Z}\left[ \text{Re}G(\epsilon ,n)\text{Re}G(\epsilon
,n+p)\right. - \\
& \left. \text{Im}G(\epsilon ,n)\text{Im}G(\epsilon ,n+p)\right] Z(n,p).
\end{split}
\label{Q}
\end{equation}%
Interlayer conductivity
\begin{equation}
\sigma _{zz}\left( T\right) =\frac{1}{2}\sum_{s=\pm 1}\int d\varepsilon %
\left[ -n_{F}^{\prime }(\varepsilon )\right] \sigma _{zz}\left( \varepsilon
\right) ,  \label{sT}
\end{equation}%
where%
\begin{equation}
\sigma _{zz}\left( \varepsilon \right) =\left( Cl-Q\right)
\,2e^{2}t_{z}^{2}d/\pi \hbar \equiv \sigma _{zz}^{0}I_{1},  \label{1}
\end{equation}%
\begin{equation}
\frac{I_{1}}{\Gamma _{0}\hbar \omega _{c}}=\frac{2}{\pi }\sum_{n,p\in
Z}Z(n,p)\text{Im}G(\varepsilon ,n)\text{Im}G(\varepsilon ,n+p),  \label{I1}
\end{equation}%
and the interlayer conductivity in the absence of magnetic field $\sigma
_{zz}^{0}=2e^{2}\tau _{0}m^{\ast }t_{z}^{2}d/\pi \hbar
^{4}=2e^{2}g_{LL}t_{z}^{2}d/\hbar ^{2}\omega _{c}\Gamma _{0}$.

When many Landau levels (LL) are filled, i.e. at $n\sim n_{LL}^{F}\equiv
\left\lceil \mu /\hbar \omega _{c}\right\rceil \gg 1$, one can use the
asymptotics of Laguerre polynomials,
\begin{eqnarray}
L_{n}^{\alpha }(z) &\approx &\frac{\Gamma (\alpha +n+1)}{n!}\left( \left( n+%
\frac{\alpha +1}{2}\right) z\right) ^{-\frac{\alpha }{2}}  \label{LnA} \\
&&\times \exp \left( \frac{z}{2}\right) J_{\alpha }\left( 2\sqrt{\left( n+%
\frac{\alpha +1}{2}\right) z}\right) ,  \notag
\end{eqnarray}%
which gives at$\,0\leq p\ll n$
\begin{equation}
Z(n,p)\approx J_{p}^{2}\left( \sqrt{2n+1}ql_{H}\right) .  \label{Z1}
\end{equation}%
Eq. (\ref{Z1}) can be further simplified using $\sqrt{2n_{LL}^{F}+1}%
ql_{H}\approx k_{F}d\tan \theta $ and that $Z(n,p)$ has a weak dependence on
$n$:
\begin{equation}
Z(n,p)\approx Z(n_{LL}^{F},p)\approx J_{p}^{2}\left( k_{F}d\tan \theta
\right) .  \label{Z2}
\end{equation}

It is often convenient to cast the sum over LL number $n$ into a sum over
harmonics using the Poisson summation formula:
\begin{equation}
\sum_{0}^{\infty }f(n)=\int_{0}^{\infty }f(n)dn+\sum_{k=-\infty }^{\infty
}\int_{0}^{\infty }f(n)\exp (2\pi ikn)dn.  \label{Poisson}
\end{equation}%
One can show that the zero-harmonic $k=0$ of the quantum part $Q$ of
interlayer conductivity $\sigma _{zz}$ in Eq. (\ref{Q}) is almost zero,
provided the dependence on $n$ of $Z(n,p)$ is much weaker than the
dependence of $G(\epsilon _{F},n)$, which is valid for $n\gg 1$. Then $%
Z(n,p)\approx J_{p}^{2}\left( k_{F}d\tan \theta \right) $ can be factored
out from the sum over $n$ in Eq. (\ref{Q}). Substituting Eq. (\ref{G1}) to
Eq. (\ref{Q}) and applying the Poisson summation formula (\ref{Poisson}) we
obtain for $k=0$:
\begin{gather}
\bar{Q}\approx \sum_{p\in Z}\frac{J_{p}^{2}\left( k_{F}d\tan \theta \right)
}{\left( \hbar \omega _{c}\right) ^{2}}\times  \label{Qm} \\
\int_{-\infty }^{\infty }\frac{dn\left[ \left( n-u\right) \left(
n+p-u\right) -v^{2}\right] }{\left[ \left( n-u\right) ^{2}+v^{2}\right] %
\left[ \left( n+p-u\right) ^{2}+v^{2}\right] }=0,  \notag
\end{gather}%
where $u\equiv \left[ \varepsilon -\text{Re}\Sigma \left( \varepsilon
\right) \right] /\hbar \omega _{c}-1/2$ and $v\equiv $Im$\Sigma \left(
\varepsilon \right) /\hbar \omega _{c}$. The integral over $n$ in Eq. (\ref%
{Qm}) is zero for each $p$, because the residues in the poles at $n=u+iv$
and $n=u-p+iv$\ cancel each other for each $p\neq 0$, while at $p=0$ the
residue is zero, which can be checked by a direct calculation. Hence, $\bar{Q%
}\approx 0$. One can show, taking the dependence $Z(n,p)$ into account, that
$\bar{Q}$ is smaller than the classical part $Cl$ by a factor $\sim
p\,dZ(n,p)/dn\sim p/n\ll 1$. Note that this statement does not depend on the
LL shape, because Eq. (\ref{Qm}) is valid for arbitrary $\Sigma \left(
\varepsilon \right) $.

\section{Results and discussion}

In a strong magnetic field, $\hbar \omega _{c}\gg \Gamma $, the details of
AMRO essentially depend on the LL shape, determined by the Green's function $%
G(\varepsilon ,n)$. Therefore, below we consider the Lorentzian and Gaussian
LL shapes separately and compare the results.

\subsection{AMRO for different LL shapes}

For Lorentzian LL shape $|\text{Im}\Sigma \left( \varepsilon \right)| =\Gamma =\Gamma
\left( B\right) $ in Eq. (\ref{G1}) is independent of $\varepsilon $. This
approximation is equivalent to that in Ref. \cite{WIPRB2011}, and for the
monotonic part $\bar{\sigma}_{zz}^{L}$ of interlayer conductivity one
confirms Eq. (\ref{sa}) with the renormalized $\sigma _{zz}^{0}$ and $\tau
=\hbar /2\Gamma $ according to Eqs. (\ref{s0R}) and (\ref{tau0R}). The
calculation of AMRO in the presence of MQO for the Lorentzian LL shape can
be performed rather simply (see Appexdix III). Combining Eqs. (\ref{sT}), (%
\ref{1}) and (\ref{3})-(\ref{Ap}) one obtains
\begin{equation}
\frac{\sigma _{zz}^{L}}{\sigma _{zz}^{0}}=\frac{\Gamma _{0}}{2\Gamma }%
\sum_{s=\pm 1}\int d\varepsilon \left[ -n_{F}^{\prime }(\varepsilon )\right]
\sum_{p=-\infty }^{\infty }A_{p}\left[ J_{p}\left( \kappa \right) %
\right] ^{2}  \label{s1L}
\end{equation}%
where $A_{0}$ and $A_{p}$ are given by Eqs. (\ref{A0}) and (\ref{Ap}).
Substituting Eq. (\ref{Iof}) to Eq. (\ref{sT}),(\ref{1}) and performing the
standard integration over $\epsilon $ one can express the result for the
angular dependence of $\sigma _{zz}^{L}$ in the presence of MQO as a
harmonic series:
\begin{gather}
\frac{\sigma _{zz}^{L}}{\sigma _{zz}^{0}}=\frac{\Gamma _{0}}{\Gamma }%
\sum_{k=-\infty }^{\infty }\left( -1\right) ^{k}\exp \left( \frac{2\pi
ik\epsilon _{F}}{\hbar \omega _{c}}\right) R_{D}\left( k\right) R_{T}\left(
k\right)   \notag \\
\times R_{S}\left( k\right) \,\left\{ \left[ J_{0}\left( \kappa \right) %
\right] ^{2}\left( 1+\frac{\pi |k|}{\omega _{c}\tau }\right)
+\sum_{p=1}^{\infty }\frac{2\left[ J_{p}\left( \kappa \right) \right] ^{2}}{%
1+\left( p\omega _{c}\tau \right) ^{2}}\right\} ,  \label{sLk}
\end{gather}%
where the Dingle factor
\begin{equation*}
R_{D}\left( k\right) =\exp \left( \frac{-\pi |k|}{\omega _{c}\tau }\right) ,
\end{equation*}%
the temperature damping factor
\begin{equation*}
R_{T}=\frac{2\pi ^{2}k_{B}T/\hbar \omega _{c}}{\sinh \left( 2\pi
^{2}k_{B}T/\hbar \omega _{c}\right) }
\end{equation*}%
and the spin factor is given by\cite{Shoenberg}%
\begin{equation}
R_{S}\left( k\right) =\cos \left( \frac{\pi gk~m^{\ast }}{2m_{e}\cos \theta }%
\right) ,  \label{Rs}
\end{equation}%
where $m^{\ast }$ and $m_{e}$ are the effective and free electron masses and
the g-factor $g\approx 2$ unless the spin-orbit or e-e interaction is
strong. The spin factor in Eq. (\ref{Rs}) also has a strong angular
dependence, giving the so-called spin-zeros angular dependence of MQO
amplitudes. The angular dependence of MQO amplitudes is given by a product
of two factors in the second line of Eq. (\ref{sLk}): $R_{S}\left( k\right) $
and the AMRO factor in the curly brackets. Hence, the traditional fitting of
the experimentally observed angular dependence of MQO amplitudes by the
spin-zero factor only is not correct. The extra factor $\left( 1+\pi
|k|/\omega _{c}\tau \right) $, multiplying $\left[ J_{0}\left( \kappa \right) %
\right] ^{2}$ in the second line of Eq. (\ref{sLk}), enhances the AMRO
amplitude of MQO as compared to the AMRO of monotonic part of MR. At $\omega
_{c}\tau \lesssim 1$ this extra factor $\left( 1+\pi |k|/\omega _{c}\tau
\right) \gg 1$, and even the ratio $\tilde{\sigma}_{zz}^{L}/\bar{\sigma}%
_{zz}^{L}$ of the oscillating and monotonic parts of conductivity has more
complicated angular dependence than just given by the spin factor $R_{S}$.

In 2D and strongly anisotropic quasi-2D layered compounds in strong magnetic
field, when $\omega _{c}\tau \gg 1$, the LL shape is not Lorentzian.\cite%
{Ando,Ando3,Brezin,Imp,Burmi,Marihin1989,KukushkinUFN1988,RMP2012,WIPRB2012}
For a physically reasonable white-noise or Gaussian correlator of the
disorder potential $U\left( \mathbf{r}\right) $, $\left\langle U\left(
\mathbf{0}\right) U\left( \mathbf{r}\right) \right\rangle \propto \exp
\left( -r^{2}/2d^{2}\right) $, theory predicts the Gaussian LL shape of the
Landau levels (for reviews see, e.g., Refs. \cite{KukushkinUFN1988} and \cite%
{RMP2012}):
\begin{equation}
\left\vert \text{Im}G(\epsilon ,n)\right\vert =\left( \sqrt{\pi }/\Gamma
\right) \exp \left[ -(\epsilon -\epsilon _{n})^{2}/\Gamma ^{2}\right] .
\label{Gauss}
\end{equation}%
At $\hbar \omega _{c}\gg \Gamma ,T$, when the LLs have Gaussian shape, only
few LLs at the Fermi level contribute to conductivity, because $\left\vert
\text{Im}G(\varepsilon =\mu ,n)\right\vert $ is negligibly small for any $%
\left\vert n-n_{LL}^{F}\right\vert ,\left\vert p\right\vert \geq 2$. Then,
in the sum over $p$ and $n$ in Eq. (\ref{I1}) one may keep only three LLs: $%
n=n_{LL}^{F}$ or $n=n_{LL}^{F}\pm 1$, and $p=0,\pm 1$:
\begin{equation}
\frac{I_{1}}{\Gamma _{0}\hbar \omega _{c}}=\frac{2}{\pi }%
\sum_{n-n_{LL}^{F},p\,=0,\pm 1}Z(n,p)\text{Im}G(\varepsilon ,n)\text{Im}%
G(\varepsilon ,n+p).  \label{s1m1}
\end{equation}%
The LL with $n=n_{LL}^{F}\pm 1$ also contain a small factor $\left\vert
\text{Im}G(\mu ,n\pm 1)\right\vert $ at $\hbar \omega _{c}\gg \Gamma ,T$.
However, they cannot be completely neglected. First, the terms $p\neq 0$ are
responsible for the damping of AMRO. Without these terms the interlayer
conductivity, given by Eqs. (\ref{sT})-(\ref{I1}),(\ref{Z2}), would be
strictly zero in the Yamaji angles. Second, when $\varepsilon /\hbar \omega
_{c}$ is integer, Im$G(\varepsilon ,n_{LL}^{F}+1)=$Im$G(\varepsilon
,n_{LL}^{F})$, and the terms $n=n_{LL}^{F},p=1$ and $n=n_{LL}^{F}+1,p=0,-1$
give the same contribution as the term $n=n_{LL}^{F},p=0$. At higher
temperature $T>\hbar \omega $ one has to keep several terms in the sum over $%
n$ but not over $p$ in Eq. (\ref{I1}), which at $\mu \gg T\gtrsim \hbar
\omega $ only very slightly affects AMRO.

\begin{figure}[tb]
\includegraphics[width=0.49\textwidth]{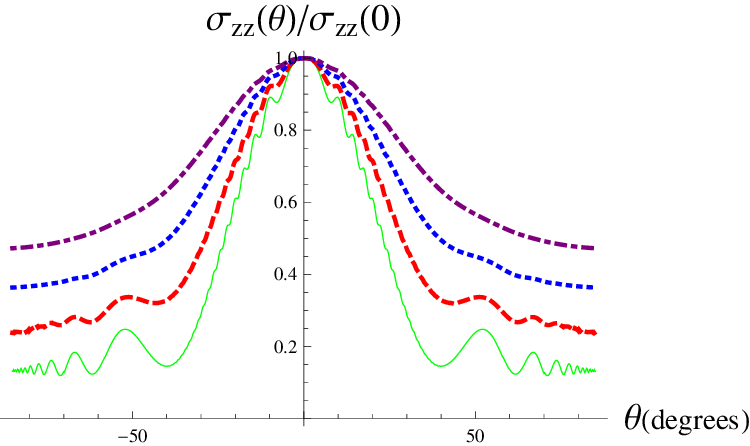}
\caption{(Color online) The angular dependence of normalized interlayer
conductivity, calculated using Eqs. (\protect\ref{sT})-(\protect\ref{I1})
for the Lorenztian LL shape with four different values of $\Gamma _{0}=\hbar
/2\protect\tau $: $\Gamma _{0}=0.5K$ (thin solid green curve), $\Gamma
_{0}=1.5K$ (dashed red curve), $\Gamma _{0}=3.0K$ (dotted blue curve), and $%
\Gamma _{0}=5.0K$ (dash-dotted purple curve). The other parameters are $%
k_{F}d=3,$ $\protect\mu =605K,$ $T=3K$, and $B_{0}\approx 11.6T$, which for
cyclotron mass $m^{\ast }=m_{e}$ and for $\protect\theta =0$ corresponds to $%
\hbar \protect\omega _{c}=10K$. }
\label{FigL1}
\end{figure}

\begin{figure}[tb]
\includegraphics[width=0.49\textwidth]{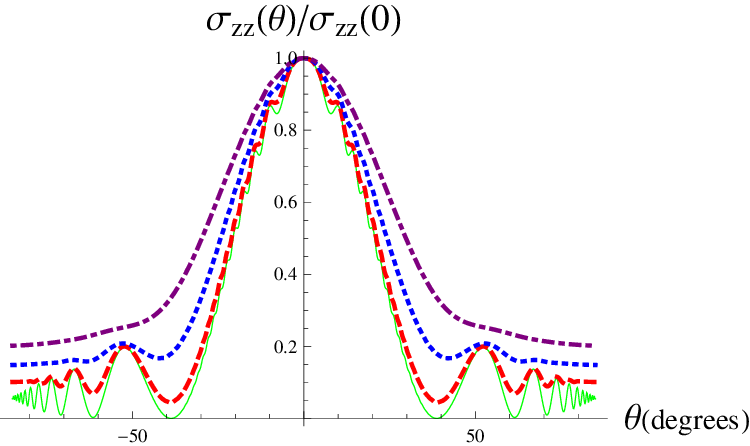}
\caption{(Color online) The same as in Fig. \protect\ref{FigL1} but for
Gaussian LL shape. The AMRO are much stronger, and the saturation value of $%
\protect\sigma _{zz}$ at $\protect\theta \rightarrow \pm 90^{\circ }$\ is
considerably smaler than for the Lorentzian LL shape.}
\label{FigG1}
\end{figure}

In Figs. \ref{FigL1} and \ref{FigG1} we plot the calculated AMRO for the
Lorentzian and Gaussian LL shapes for four different values of $\Gamma
_{0}=\hbar /2\tau =0.5K,1.5K,3.0K$ and $5.0K$, corresponding to $\omega
_{c0}\tau =10,3.33,1.67$ and $1.0$ respectively at $\theta =0$. Comparison
of Figs. \ref{FigL1} and \ref{FigG1} shows that the same value of $\Gamma
_{0}$ suppresses AMRO much stronger for the Lorentzian LL shape than for the
Gaussian. In particular, at finite $\Gamma \lesssim \hbar \omega $ the
minima of conductivity at the Yamaji angles are much deeper for the Gaussian
LL shape than for Lorentzian. Neglecting this may lead to the incorrect
determination of $\omega _{c}\tau $ from the experimental data on AMRO
amplitude.

At $\Gamma \ll \hbar \omega $, Eq. (\ref{s1m1}) gives exponentially small
values of $\sigma _{zz}^{G}$ $\sim \sigma _{zz}^{0}\exp \left[ -\left( \hbar
\omega _{c}/2\Gamma \right) ^{2}\right] $ in the Yamaji angles. Besides
finite LL broadening $\Gamma $, MR in the Yamaji maxima is limited by the
additional "incoherent" mechanisms of interlayer transport, such as the
interlayer hopping via resonance impurities\cite%
{Abrikosov1999,Incoh2009,Maslov} and dislocations, or the boson-assisted
tunneling.\cite{Lundin2003,Ho} Approximately, the contribution of the
incoherent channels to $\sigma _{zz}$ does not depend on the tilt angle $%
\theta $ of magnetic field and gives a constant upward shift of the curves
in Figs. \ref{FigL1} and \ref{FigG1}.

\subsection{High tilt angle}

From Figs. \ref{FigL1} and \ref{FigG1} one observes that not only the AMRO
amplitude but also the ratio $\sigma _{zz}\left( \theta \rightarrow \pm
90^{\circ }\right) /\sigma _{zz}\left( \theta =0\right) $ depend on the LL
shape: the saturation value of $\sigma _{zz}$ at $\theta \rightarrow \pm
90^{\circ }$\ looks considerably smaller for the Gaussian LL shape than for
the Lorentzian. However, the calculated absolute values of $\sigma
_{zz}\left( \theta \rightarrow \pm 90^{\circ }\right) /\sigma _{zz}^{0}$
depend only on $\omega _{c0}\tau \cdot k_{F}d$ but not on the LL shape.
These values agree well with Eq. (10) of Ref. \cite{ShofieldCooper2000},
which predicts
\begin{equation}
\sigma _{zz}\left( \theta \rightarrow \pm 90^{\circ }\right) /\sigma
_{zz}^{0}=1/\sqrt{1+\left( k_{F}d\omega _{c0}\tau \right) ^{2}},
\label{spar}
\end{equation}%
where $\omega _{c0}=eB_{0}/m^{\ast }c$. In Ref. \cite{ShofieldCooper2000}
Eq. (\ref{spar}) was obtained in the $\tau $-approximation using the
quasi-classical electron trajectories along the well-defined 3D Fermi
surface. The $\tau $-approximation does not work in strong
perpendicular-to-layers magnetic field, but it may work properly when the
magnetic field is along the conducting layers so that $B_{z}\rightarrow 0$.
One can also expect that the LL shape is not important in the limit $\theta
\rightarrow \pm 90^{\circ }$ and $B_{z}\rightarrow 0$. To check this, we now
calculate $\sigma _{zz}\left( \theta \rightarrow \pm 90^{\circ }\right)
/\sigma _{zz}^{0}$ for the Lorentzian and Gaussian LL shapes without the use
of a 3D FS and of the $\tau $-approximation.

At high tilt angle the argument of the Bessel's functions in Eq. (\ref{Z2}) $%
\kappa \equiv k_{F}d\tan \theta \gg 1$, and one can use its asymptotic
expansion, which gives%
\begin{eqnarray}
Z(n,p) &\approx &\left( 2/\pi \kappa \right) \cos ^{2}\left( \kappa -\pi
p/2-\pi /4\right)   \notag \\
&=&\left[ 1+\cos \left( 2\kappa -\pi p-\pi /2\right) \right] /\pi \kappa .
\label{Zpar}
\end{eqnarray}%
The square brackets contain a sum of the monotonic and alternating terms as
function of $p$. At $B_{z}\rightarrow 0$, when the LL\ separation $\hbar
\omega _{c}\ll \Gamma $, the factor Im$G(\varepsilon ,n+p)$ in Eq. (\ref{I1}%
) depends very weakly on $p$, and the alternating term gives a negligible
contribution to Eq. (\ref{I1}). Substituting only a constant term from Eq. (%
\ref{Zpar}) to Eq. (\ref{I1}) gives at $\theta \rightarrow \pm 90^{\circ }$
\begin{equation}
\frac{I_{1}}{\Gamma _{0}\hbar \omega _{c}}\approx \frac{2}{\pi ^{2}\kappa }%
\sum_{n,p\in Z}\text{Im}G(\varepsilon ,n)\text{Im}G(\varepsilon ,n+p).
\end{equation}%
At $\hbar \omega _{c}\equiv \hbar eB_{z}/m^{\ast }c\ll \Gamma $ one can
replace the summations over $n$ and $p$ by the integrations. For the
Lorentzian LL shape this gives
\begin{eqnarray}
I_{1} &\approx &\int \int_{-\infty }^{\infty }\frac{dpdn~\Gamma _{0}\hbar
\omega _{c}\Gamma ^{2}\left( 2/\pi ^{2}\kappa \right) }{\left[ (\epsilon
-\epsilon _{n})^{2}+\Gamma ^{2}\right] \left[ (\epsilon -\epsilon
_{n+p})^{2}+\Gamma ^{2}\right] }  \notag \\
&=&2\Gamma _{0}/\kappa \hbar \omega _{c}=\left( \omega _{c0}\tau
k_{F}d\right) ^{-1}  \label{I1LP}
\end{eqnarray}%
in agreement with Eq. (14) of Ref. \cite{MosesMcKenzie1999}. For Gaussian LL
shape at $\theta \rightarrow \pm 90^{\circ }$\ we obtain the same result:
\begin{gather}
I_{1}\approx \frac{2\Gamma _{0}\hbar \omega _{c}}{\pi \kappa \Gamma ^{2}}%
\int_{-\infty }^{\infty }dn\exp \left[ -\frac{(\epsilon -\epsilon _{n})^{2}}{%
\Gamma ^{2}}\right] \times   \notag \\
\times \int_{-\infty }^{\infty }dp\exp \left[ -\frac{(\epsilon -\epsilon
_{n+p})^{2}}{\Gamma ^{2}}\right] =\frac{2\Gamma _{0}}{\kappa \hbar \omega
_{c}}.  \label{I1GP}
\end{gather}%
Thus, the ratio $\sigma _{zz}\left( \theta \rightarrow \pm 90^{\circ
}\right) /\sigma _{zz}^{0}=\left( \omega _{c0}\tau k_{F}d\right) ^{-1}$ is
the same for Lorentzian and Gaussian LL shapes. This result is natural,
because when $\theta \rightarrow \pm 90^{\circ }$ and $B_{z}$ is small, so
that $\Gamma \gg \hbar \omega _{c}$, the LLs are smeared and their shape is
not important. However, $\sigma _{zz}^{0}\neq \sigma _{zz}\left( \theta
=0\right) $, and $\sigma _{zz}\left( \theta =0\right) $ depends on the LL
shape. Substitution of Eqs. (\ref{Lor}) and (\ref{Gauss}) to Eq. (\ref{s1m1}%
), keeping only one term $n=n_{LL}^{F},p=0$, gives that at $T=0$ in the
maxima of MQO the value of $\sigma _{zz}\left( \theta =0\right) $ for the
Gaussian LL shape is $\pi $ times larger than for the Lorentzian for the
same $\Gamma $. Therefore, in Fig. \ref{FigG1} the ratio $\sigma _{zz}\left(
\theta \rightarrow \pm 90^{\circ }\right) /\sigma _{zz}\left( \theta
=0\right) $ is considerably smaller than in Fig. \ref{FigL1}. Since $k_{F}d$
is usually known from the AMRO\ period, and $\omega _{c0}$ (determined by
the effective mass $m^{\ast }$) is known from the MQO\ period, the
experimentally obtained ratio $\sigma _{zz}\left( \theta \rightarrow \pm
90^{\circ }\right) /\sigma _{zz}^{0}$ provides a tool to determine $\tau $
with high accuracy.\

\subsection{Spin current and the influence of spin on AMRO}

The spin current, as a key object of spintronics, attracts a great attention
for its present-day and potential applications (see, e.g., Refs. \cite%
{SpinCurrentBook,SpintronicsRMP} for reviews). In our system, the non-zero
spin current conductivity $s_{zz}\equiv \sigma _{zz\uparrow }-\sigma
_{zz\downarrow }$ appears because the electrons with opposite spin
orientations give nonequal contributions to $\sigma _{zz}$. The Fermi energy
of spin up and down electrons differs by the Zeeman energy $g\mu _{B}B$,
giving different phase of MQO, which leads to the MQO of the spin current.
The MQO amplitudes of the spin-current conductivity $s_{zz}$ and of the
usual charge conductivity $\sigma _{zz}$ have completely different angular
dependence. For the Lorentzian LL shape and neglecting the Zeeman splitting
of the Fermi momentum $k_{F}$ in the argument of the Bessel's functions, the
MQO amplitudes of the spin-current are given by the second line of Eq. (\ref%
{sLk}) with the replacement $R_{S}\left( k\right) \rightarrow
R_{S}^{spin}\left( k\right) $, instead of Eq. (\ref{Rs}) given by
\begin{equation}
R_{S}^{spin}\left( k\right) =\sin \left( \pi gk~m^{\ast }/2m_{e}\cos \theta
\right) .  \label{RsSpin}
\end{equation}%
The spin-current MQO are damped by temperature and disorder, similar to the
usual MQO. Hence, the measurement of the spin-current MQO is not simpler
than the measurement of the usual MQO, and this measurement is useful only
if it gives any additional information about a compound, which cannot be
extracted from the MQO of $\sigma _{zz}$. For example, if the $g$-factor
cannot be reliably extracted from the spin-zero angles of the usual MQO in
the available range of tilt angles, or if these spin-zero angles casually
coincide with the Yamaji angles. The non-sinusoidal shape of MQO and the
interplay of MQO and AMRO make the angular dependence of the harmonic
amplitudes of both $\sigma _{zz}$ and $s_{zz}$\ even more complicated than
just given by Eqs. (\ref{Rs}) and (\ref{RsSpin}). For example, one can
observe only minima of the conductivity harmonic amplitudes instead of
strict spin zeros, given by Eqs. (\ref{Rs}) and (\ref{RsSpin}).

\begin{figure}[tb]
\includegraphics[width=0.49\textwidth]{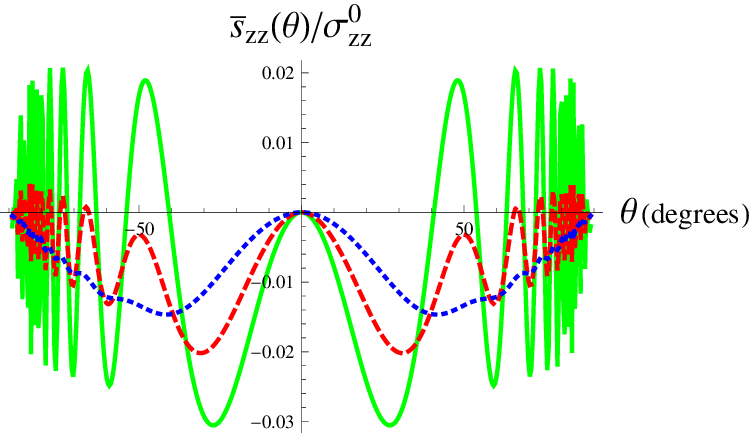}
\caption{(Color online) The angular dependence of the monotonic part of the
spin-current conductivity $\bar{s}_{zz}/\protect\sigma _{zz}^{0}$,
calculated from Eq. (\protect\ref{sLk}) for $k=0$ for three different values
of $\Gamma $, independent of $B_{z}$ and corresponding to $\protect\omega %
_{c}\protect\tau =10$ (solid green line), $1.0$ (dashed red line) and $0.5$
(dotted blue line). }
\label{FigSpinLAMRO}
\end{figure}

The monotonic part of spin current appears mainly because of the slightly
different angular dependence of the contributions to conductivity from
electrons with different spin. The difference of Fermi momenta for spin up
and down electrons, originating from the Zeeman energy splitting $g\mu _{B}H$%
, leads to the difference $\delta \kappa $ in the argument of the Bessel's
functions in Eq. (\ref{Z2}):
\begin{equation*}
\delta \kappa =g\mu _{B}B_{0}\tan \theta \,d/\hbar v_{F}\approx 2\mu
_{B}B_{0}m^{\ast }d\tan (\theta )/(\hbar ^{2}k_{F}).
\end{equation*}%
The monotonic part $\bar{s}_{zz}$ of the spin-current conductivity,
determined as the difference between the monotonic parts of conductivities
with spin up and down as $\bar{s}_{zz}\approx \bar{\sigma}_{zz}(\kappa
+\delta \kappa )-\bar{\sigma}_{zz}(\kappa )$, for the Lorentzian LL shape in
the first order in $\kappa \ll 1$ is given by
\begin{eqnarray}
\frac{\bar{s}_{zz}}{\sigma _{zz}^{0}} &\approx &\left\{ 2J_{0}\left( \kappa
\right) J_{0}^{\prime }\left( \kappa \right) +4\sum_{\nu =1}^{\infty }\frac{%
[J_{\nu }\left( \kappa \right) J_{\nu }^{\prime }\left( \kappa \right) ]}{%
1+\left( \nu \omega _{c}\tau \right) ^{2}}\right\} \delta \kappa   \notag \\
&=&-\delta \kappa \sum_{\nu =-\infty }^{\infty }\frac{J_{\nu }\left( \kappa
\right) [J_{\nu +1}\left( \kappa \right) -J_{\nu -1}\left( \kappa \right) ]}{%
1+\left( \nu \omega _{c}\tau \right) ^{2}},  \label{sAv}
\end{eqnarray}%
where we have applied $2J_{\nu }^{\prime }\left( \kappa \right)
=J_{\nu -1}\left( \kappa \right) -J_{\nu +1}\left( \kappa \right)
$. In a field $ B_{0}=10T$ and for the parameters
$d=20\mathring{A}$, $k_{F}=0.14\mathring{A}^{-1}$ and $m^{\ast
}\approx 2m_{e}$, corresponding to the organic metal $\alpha
$-(BEDT-TTF)$_{2}$KHg(SCN)$_{4}$ (see Ref. \cite{MarkPRL2006}),
$\delta \kappa \approx 0.1\tan \theta $ is not negligible. For
these parameters, in Fig. \ref{FigSpinLAMRO} we plot the angular
dependence of $\bar{s}_{zz}/\sigma _{zz}^{0}$, calculated without
expansion in $\delta \kappa $, i.e. from Eq. (\ref{sLk}) for $k=0$
(neglecting the MQO), for three different values of $\Gamma $,
independent of $B_{z}$ and corresponding to $\omega _{c}\tau =10$
(solid green line)$,1$ (dashed red line) and $0.5$ (dotted blue
line). We also checked that the first-order expansion in $\delta
\kappa \approx 0.1\tan \theta $, given by Eq. (\ref{sAv}), works
very well for $\left\vert \theta \right\vert <86^{\circ }$.

In the Yamaji angles $\sigma _{zz}\left( \theta \right) /\sigma _{zz}^{0}\ll
1$, and the spin current for these angles can be comparable to the charge
current, being also considerably smaller than for other angles at $\omega
_{c}\tau \gg 1$. Note, that at $\omega _{c}\tau \gg 1$ the monotonic part of
spin current changes sign in the proximity of the Yamaji angles from "-" to
"+", and it changes its sign back in the extrema of $\sigma _{zz}\left(
\theta \right) $. In heterostructures the spin current can be considerably
larger than shown in Fig. \ref{FigSpinLAMRO}, because of a larger value of $%
\delta \kappa $, which is proportional to the interlayer distance $d$.

\section{Conclusions}

We have presented the quantum-mechanical calculations of the angular
dependence of interlayer magnetoresistance in quasi-2D layered metals. The
previous calculations of AMRO usually neglected the magnetic quantum
oscillation,\cite{Kur,MosesMcKenzie1999} or even used the semiclassical
Boltzmann transport equation in the constant-$\tau $ approximation.\cite%
{Yagi1990,Mark92,Singleton2001,GrigAMRO2010} However, even if MQO are not
seen, being damped by temperature or long-range disorder, they strongly
influence the interlayer conductivity and its angular dependence in a strong
magnetic field, when $\hbar \omega _{c}\gg t_{z},\Gamma $.\cite%
{WIPRB2011,WIJETP2011,WIFNT2011,GrigPRB2013} In the present study
we take MQO into account from the beginning and consider the
influence of MQO on AMRO. Our calculation is applicable for
various shapes of the Landau levels, thus generalizing the
calculation in Refs. \cite{Kur,MosesMcKenzie1999,WIPRB2011}. This
is important, because when the interlayer transfer integral
$t_{z}$ is less than the LL separation $\hbar \omega _{c}$, the LL
shape is not Lorentzian\cite%
{Ando,Ando3,Brezin,Imp,Burmi,Marihin1989,KukushkinUFN1988,WIPRB2012}
In addition, we take into account the so-called "quantum term" in
the magnetoresistance,\cite{ShubCondMat1,Shub,ChampelMineev} also
originating from MQO and neglected in the previous
studies.\cite{Kur,MosesMcKenzie1999,WIPRB2011}

Our calculation shows that the LL shape is important for the angular
dependence of magnetoresistance. In particular, the AMRO amplitude is much
stronger for the Gaussian LL shape than for the Lorentzian (compare Figs. %
\ref{FigL1} and \ref{FigG1}). The ratio $\sigma _{zz}\left( \theta
=0\right) /\sigma _{zz}\left( \theta \rightarrow \pm 90^{\circ
}\right) $ is also several times larger for the Gaussian LL shape.
For the Lorentzian LL shape the angular dependence of interlayer
conductivity is given by Eqs. (\ref{s1L}),(\ref{A0}),(\ref{Ap}) or
by Eqs. (\ref{sLk})-(\ref{Rs}), which combine MQO and
AMRO.\cite{CommentChampel} For arbitrary LL shape one can use Eqs.
(\ref{sT})-(\ref{I1}) with $Z(n,p)$ given by Eq. (\ref{Z}) or by
Eqs. (\ref{Z1}) or (\ref{Z2}). In the high-field limit one can
apply Eq. (\ref{s1m1}) instead of Eq. (\ref{I1}) to calculate
$\sigma _{zz}\left( \theta \right) $.

We also estimated the spin current, which appears because of AMRO. For
typical parameters of the organic metal $\alpha $-(BEDT-TTF)$_{2}$KHg(SCN)$%
_{4}$ and in the field $B \sim 10T$ the spin current is about 2\%
of the zero-field charge current (see Fig. \ref{FigSpinLAMRO}),
but it may almost reach the charge current for special tilt angles
of magnetic field. In heterostructures the spin current can be
considerably larger. The angular oscillations of the spin current
are stronger and shifted by the phase $\sim \pi /2$ as compared to
the usual charge-current AMRO.
\medskip

The work was supported by the Russian Foundation for Basic Research.

\appendix

\section{Classical part of conductivity}

Substituting Eq. (\ref{GNC}) to the first line of Eq. (\ref{Cl}) one obtains
\begin{eqnarray*}
Cl &=&\int dy_{2}dy_{1}dx_{2}dx_{1}\cos \left[ q(y_{2}-y_{1})\right] \\
&&\times \sum_{p,n,k_{y},k_{y}^{\prime }}\Psi _{n,k_{y}^{\prime }}^{\ast
}(r_{1})\Psi _{n,k_{y}^{\prime }}(r_{2})G(\epsilon ,n) \\
&&\times \Psi _{n+p,k_{y}}^{\ast }(r_{2})\Psi _{n+p,k_{y}}(r_{1})G^{\ast
}(\epsilon ,n+p),
\end{eqnarray*}%
where the wave functions are given by Eqs. (\ref{Psi}) and (\ref{Gg}).
Integration over $y_{2}$, $y_{1}$ (in a unit square) gives:
\begin{align}
& Cl=4\pi ^{2}\text{Re}\int dx_{2}dx_{1}\sum_{p,n,k_{y},k_{y}^{\prime }}\Psi
_{n+p}^{\ast }(x_{2}-l_{H}^{2}k_{y})  \label{Cl1} \\
& \times \Psi _{n+p}(x_{1}-l_{H}^{2}k_{y})\Psi _{n}^{\ast
}(x_{1}-l_{H}^{2}k_{y}^{\prime })\Psi _{n}(x_{2}-l_{H}^{2}k_{y}^{\prime })
\notag \\
& \times G^{\ast }(\epsilon ,n+p)G(\epsilon ,n)\delta (k_{y}+q-k_{y}^{\prime
}).  \notag
\end{align}%
Summation over $k_{y}^{\prime }$ cancels $\delta $-function. Then we use the
identity:
\begin{eqnarray}
&&\int_{-\infty }^{\infty }dxe^{-c^{2}x^{2}}H_{n}(a+cx)H_{n+p}(b+cx)
\label{Exp:1} \\
&=&\frac{2^{n}\sqrt{\pi }n!b^{p}}{c}L_{n}^{p}(-2ab),\,0\leq p.  \notag
\end{eqnarray}%
Using Eqs. (\ref{Exp:1}) and (\ref{Gg}) one may get:%
\begin{eqnarray}
&&\int_{-\infty }^{\infty }dx\Psi _{n+p}(x-l_{H}^{2}k_{y})\Psi
_{n}(x-l_{H}^{2}(k_{y}+q))=  \label{IPsi} \\
&&\exp \left( -\frac{(ql_{H})^{2}}{4}\right) \left( \frac{ql_{H}}{\sqrt{2}}%
\right) ^{p}L_{n}^{p}\left( \frac{(ql_{H})^{2}}{2}\right) \sqrt{\frac{n!}{%
(n+p)!}}.  \notag
\end{eqnarray}%
The integration over $x_{1},x_{2}$ in Eq. (\ref{Cl1}) is performed using Eq.
(\ref{IPsi}). Then, making the summation over $k_{y}$, which just gives the
LL degeneracy $g_{LL}=1/2\pi l_{H}^{2}=eB_{z}/2\pi \hbar c$, we obtain Eq. (%
\ref{Cl}).

\section{Quantum Part of conductivity}

Substituting Eq. (\ref{GNC}) to the first line of Eq. (\ref{Q}) gives%
\begin{eqnarray*}
Q &=&\int dy_{2}dy_{1}dx_{2}dx_{1}\exp \left[ iq(y_{2}-y_{1})\right] \\
&&\times \sum_{p,n,k_{y},k_{y}^{\prime }}\Psi _{n,k_{y}^{\prime }}^{\ast
}(r_{1})\Psi _{n,k_{y}^{\prime }}(r_{2})G(\epsilon ,n) \\
&&\times \Psi _{n+p,k_{y}}^{\ast }(r_{2})\Psi _{n+p,k_{y}}(r_{1})G(\epsilon
,n+p),
\end{eqnarray*}%
which after the substitution of Eq. (\ref{Psi}) and integration over $%
y_{1},y_{2}$ becomes
\begin{align*}
& Q=4\pi ^{2}\text{Re}\int dx_{2}dx_{1}\sum_{p,n,k_{y},k_{y}^{\prime }}\Psi
_{n+p}^{\ast }(x_{2}-l_{H}^{2}k_{y}) \\
& \times \Psi _{n+p}(x_{1}-l_{H}^{2}k_{y})\Psi _{n}^{\ast
}(x_{1}-l_{H}^{2}k_{y}^{\prime })\Psi _{n}(x_{2}-l_{H}^{2}k_{y}^{\prime }) \\
& \times G(\epsilon ,n+p)G(\epsilon ,n)\delta (k_{y}+q-k_{y}^{\prime }).
\end{align*}%
The integration over $x_{1},x_{2}$ is similar to that in Eq.
(\ref{Cl1}) and can be easily done using Eq. (\ref{IPsi}).
Summation over $k_{y}$ gives the LL degeneracy. Performing these
integrations we obtain Eq. (\ref{Q}).

\section{Harmonic expansion of interlayer conductivity for the Lorentzian LL
shape}

For the Lorentzian LL shape one can put $|\text{Im}\Sigma \left( \varepsilon \right)|
=\Gamma =const$ in Eq. (\ref{G1}). Then the imaginary part of the electron
Green's function
\begin{equation}
\text{Im}G(\epsilon ,n)=\Gamma /\left[ (\epsilon -\epsilon _{n})^{2}+\Gamma
^{2}\right] .  \label{Lor}
\end{equation}%
Substituting this and Eq. (\ref{Z2}) to Eq. (\ref{I1}) one obtains
\begin{equation}
I_{1}=\sum_{n,p\in Z}\frac{\left( 2/\pi \right) \hbar \omega _{c}\Gamma
_{0}\Gamma ^{2}\left[ J_{p}\left( \kappa \right) \right] ^{2}}{\left[
(\epsilon -\epsilon _{n+p})^{2}+\Gamma ^{2}\right] \left[ (\epsilon
-\epsilon _{n})^{2}+\Gamma ^{2}\right] }.  \label{3}
\end{equation}%
The low limit of the summation over $n$ in Eq. (\ref{3}) can be extended to $%
-\infty $, because many LL are filled but only few LLs at the Fermi level $%
E_{F}$, i.e. with LL number $n\approx E_{F}/\hbar \omega _{c}\gg 1$,
contribute considerably to conductivity. Eq. (\ref{Z2}) and, hence, Eq. (\ref%
{3}) are valid at $n\gg 1$, and $\left\vert p\right\vert \ll n$. The
summation over over $n$ in Eq. (\ref{3}) can be easily performed using the
identities:%
\begin{equation}
\begin{split}
A_{0} \equiv \sum_{n\in Z}\frac{\left( 2/\pi \right) \hbar \omega _{c}\Gamma
^{3}}{\left[ \left( \epsilon -\hbar \omega _{c}\left( n+1/2\right) \right)
^{2}+\Gamma ^{2}\right] ^{2}} \\
=\frac{\sinh \left( 2\pi \Gamma /\hbar \omega _{c}\right) }{\cosh \left(
2\pi \Gamma /\hbar \omega _{c}\right) +\cos \left( 2\pi \epsilon /\hbar
\omega _{c}\right) }- \\
-\frac{2\pi \Gamma }{\hbar \omega _{c}}\frac{1+\cos \left( \frac{2\pi
\epsilon }{\hbar \omega _{c}}\right) \cosh \left( \frac{2\pi \Gamma }{\hbar
\omega _{c}}\right) }{\left( \cosh \left( \frac{2\pi \Gamma }{\hbar \omega
_{c}}\right) +\cos \left( \frac{2\pi \epsilon }{\hbar \omega _{c}}\right)
\right) ^{2}}
\end{split}
\label{A0}
\end{equation}%
in agreement with Eq. (23) of Ref. \cite{ChampelMineev}, and for $p\neq 0$%
\begin{eqnarray}
A_{p} &\equiv &\sum_{n\in Z}\frac{\left( 2/\pi \right) \hbar \omega
_{c}\Gamma ^{3}}{\left[ \left( \epsilon -\hbar \omega _{c}\left(
n+1/2\right) \right) ^{2}+\Gamma ^{2}\right] }  \notag \\
&&\times \frac{1}{\left[ \left( \epsilon -\hbar \omega _{c}\left(
n+p+1/2\right) \right) ^{2}+\Gamma ^{2}\right] }  \label{Ap} \\
&=&\frac{\sinh (2\pi \Gamma /\hbar \omega _{c})/\left[ 1+p^{2}\left( \hbar
\omega _{c}/2\Gamma \right) ^{2}\right] }{\cos (2\pi \epsilon /\hbar \omega
_{c})+\cosh (2\pi \Gamma /\hbar \omega _{c})}.  \notag
\end{eqnarray}

However, we are mainly interest in the monotonic part and in the harmonic
expansion of MQO, which can be obtained using the Poisson summation formula (%
\ref{Poisson}). The monotonic part of Eq. (\ref{3}) is%
\begin{eqnarray}
\bar{I}_{1} &=&\sum_{p\in Z}\int_{-\infty }^{\infty }\frac{dn~~\left( 2/\pi
\right) \hbar \omega _{c}\Gamma _{0}\Gamma ^{2}\left[ J_{p}\left( \kappa
\right) \right] ^{2}}{\left[ (\epsilon -\epsilon _{n+p})^{2}+\Gamma ^{2}%
\right] \left[ (\epsilon -\epsilon _{n})^{2}+\Gamma ^{2}\right] }  \notag \\
&=&\frac{\Gamma _{0}}{\Gamma }\sum_{p\in Z}\frac{\,\left[ J_{p}\left( \kappa
\right) \right] ^{2}}{1+p^{2}\left( \hbar \omega _{c}/2\Gamma \right) ^{2}}
\label{Im}
\end{eqnarray}%
in agreement with Eq. (\ref{sa}) with the renormalized $\sigma _{zz}^{0}$
and $\tau =\hbar /2\Gamma $ according to Eqs. (\ref{s0R}) and (\ref{tau0R}).
The harmonic expansion of $I_{1}$ is%
\begin{equation}
\tilde{I}_{1}=\sum_{k,p\in Z}\int_{-\infty }^{\infty }dn\frac{\left( 2/\pi
\right) \hbar \omega _{c}\Gamma _{0}\Gamma ^{2}\left[ J_{p}\left( \kappa
\right) \right] ^{2}\exp (2\pi ikn)}{\left[ (\epsilon -\epsilon
_{n+p})^{2}+\Gamma ^{2}\right] \left[ (\epsilon -\epsilon _{n})^{2}+\Gamma
^{2}\right] }.  \label{Io}
\end{equation}%
For $p=0$ the integration over $n$ reduces to%
\begin{eqnarray}
&&\int_{-\infty }^{\infty }\frac{dn}{\pi /2}\frac{\hbar \omega _{c}\Gamma
^{3}\exp (2ikn)}{\left[ (\epsilon _{n}-\epsilon )^{2}+\Gamma ^{2}\right] ^{2}%
}  \label{I1m} \\
&=&\left( 1+\frac{2\pi |k|\Gamma }{\hbar \omega _{c}}\right) \exp \left( 2\pi %
\frac{ik\epsilon -|k|\Gamma }{\hbar \omega _{c}}\right) .  \notag
\end{eqnarray}%
For $p\neq 0$ the integral over $n$ is%
\begin{eqnarray}
&&\int_{-\infty }^{\infty }\frac{dn~\left( 2/\pi \right) \hbar \omega
_{c}\Gamma ^{3}\exp (2ikn)}{\left[ (\epsilon _{n+p}-\epsilon )^{2}+\Gamma
^{2}\right] \left[ (\epsilon _{n}-\epsilon )^{2}+\Gamma ^{2}\right] }  \notag
\\
&=&\exp \left( 2\pi \frac{ik\epsilon -|k|\Gamma }{\hbar \omega _{c}}\right)
\frac{1}{1+\left( p\hbar \omega _{c}/2\Gamma \right) ^{2}}.  \label{I1p}
\end{eqnarray}%
Substituting Eqs. (\ref{I1m}) and (\ref{I1p}) to Eq. (\ref{Io}) we obtain ($%
\tau =\hbar /2\Gamma $)%
\begin{eqnarray}
\tilde{I}_{1} &=&\frac{\Gamma _{0}}{\Gamma }\sum_{k=-\infty }^{\infty
}\left( -1\right) ^{k}\exp \left( \frac{2\pi ik\epsilon }{\hbar \omega _{c}}%
\right) \exp \left( \frac{-\pi |k|}{\omega _{c}\tau }\right) \times  \notag \\
&&\left( \left[ J_{0}\left( \kappa \right) \right] ^{2}\left( 1+\frac{\pi |k|}{%
\omega _{c}\tau }\right) +\sum_{p\in Z}\frac{\left[ J_{p}\left( \kappa
\right) \right] ^{2}}{1+\left( p\omega _{c}\tau \right) ^{2}}\right) .
\label{Iof}
\end{eqnarray}


\begin{thebibliography}{99}
\bibitem{CommentChampel} A formula for $\sigma _{zz}$, similar to Eq. (\ref{A0}) was
previously derived in Ref. \cite{ChampelMineev}, which is
applicable only for $\theta =0$.
\end{thebibliography}
\end{document}